\documentclass[letterpaper,11pt]{article}
\usepackage{amsmath}
\usepackage{amsfonts}
\usepackage{amssymb}
\usepackage{color}
\usepackage{mathrsfs}
\usepackage{latexsym}
\usepackage{physics}
\usepackage{cite}
\usepackage{graphicx}
\usepackage{float}
\usepackage{siunitx}
\title{The Effect of $\alpha$-Vacua on the Scalar and Tensor Spectral Indices: Slow-Roll Approximation}
\author{H. Bouzari Nezhad$^1$, F. Shojai$^{1,2}$,\\$^1$Department of Physics, University of Tehran,\\Tehran, Iran.\\$^2$Foundations of Physics Group, School of Physics,\\Institute for Research in Fundamental Sciences (IPM),\\Tehran, Iran.\\}
\date{}
\begin{document}

\maketitle
\begin{abstract}
Since the duration of inflation is finite, imposing the initial condition in infinite past, i.e. the Bunch-Davies vacuum, is inherently ambiguous. In this paper, we resort to the mixed states as initial condition which are called the $\alpha$-vacua and then introduce a physical momentum cutoff $\Lambda$ \cite{Danielsson:2002kx}, in which the evolution of perturbations begins. We show that the initial time $t_i$, when the initial condition is imposed, depends on the wave number of fluctuation, as it is for the time of horizon crossing, $t_q$. Then we calculate the corrections to the scalar and tensor power spectra and their corresponding spectral indices. Throughout this work, the calculation is done up to the first order in slow-roll parameters. We indicate that the leading order corrections to the spectral indices have a $q$-dependent amplitude, $2\epsilon_{f}[2\epsilon_{f}(\frac{q}{q_{f}})^{4\epsilon+4\eta}-\eta_{f}(\frac{q}{q_{f}})^{3\epsilon+\xi}]$ times a $q$-dependent oscillatory part, $\cos(\frac{2\Lambda(q/q_{f})^{\epsilon}}{H_{f}})$, where $H$, $\epsilon$, $\eta$, and $\xi$ are the Hubble and slow-roll parameters respectively, and the subscript $f$ denotes that these quantities are evaluated at the time when the first scale, $q_{f}$, satisfies the initial condition, i.e. $q=a(t_i)\Lambda$.
\end{abstract}
\section{Introduction}
\label{sec:Introduction}
The inflationary scenario solves several puzzles of the standard big bang cosmology. For a good introduction on inflation see \cite{Linde:2005ht,Liddle:2000cg} and references therein. The problems of flatness, horizon, monopoles and density perturbations have been solved by adding a very fast expanding epoch in the history of the universe \cite{Guth:1980zm,Linde:1983gd,Starobinsky:1980te,Sato:1980yn}. In addition to solving these old problems of standard cosmology, inflation also makes some predictions that can be tested with high accuracy. Likely, the most interesting prediction of inflation is the cosmic microwave background (CMB) radiation  anisotropies. In a usual inflationary paradigm, inflation is driven by a single massless scalar field, inflaton. The role of inflation is that it magnifies tiny quantum fluctuations generated in the very early age of the universe. Eventually these fluctuations grow and, due to gravitational instability, lead to structure formation in our universe. This mechanism for structure formation is in good agreement with observations \cite{Bennett:2012zja,Ade:2015lrj,Ade:2015tva}. The fluctuations leave an imprint on the CMB radiation and studies of these anisotropies can be used to test the inflationary model. Therefore the details of inflationary model written on the high energy scale, can be reflected in the large scale structures that we observe today.\\
The recent progress of the observational cosmology, has led to deeper understanding of the universe. These observations impose some constraints on the initial conditions of inflationary fluctuations and trans-Planckian physics. The cosmological observations indicate that the power spectrum of comoving curvature perturbations is nearly scale
independent and the corresponding statistical distribution is Gaussian. However, there are several key factors affecting these aspects, including multi-field dynamics, the lack of slow-roll approximation, using of non-trivial vacuum, the self-interactions of the inflaton field and some nonlinear effects of gravity. Here we are dealing with different vacuum issue which may lead to scale dependent power spectrum. The most important non-trivial choices of vacuum include the $\alpha$-vacua \cite{Danielsson:2002kx}, the coherent state \cite{Kundu:2011sg,Kundu:2013gha}, the $\alpha$-states \cite{Mottola:1984ar,Allen:1985ux}, the thermal state \cite{Bhattacharya:2005wn}, and the excited-de Sitter modes \cite{Yusofi:2014mta}.\\
It is a well-known fact that the rapid expansion of the universe, due to inflation, stretches the wavelength of the fluctuations modes and eventually the size of these modes reaches to an observable size. In some inflationary models, the number of e-foldings is larger than the minimal number required to solve the problems of standard cosmology \cite{Guth:1980zm,Linde:1983gd,Linde:2005ht}. If inflation lasts more than about 70 e-foldings, therefore the physical wavelengths of comoving scales which create the current anisotropies in CMB, were smaller than the Planck scale at the onset of inflation. So, to study these fluctuations, the effect of trans-Planckian physics cannot be ignored. This means that, it is possible that some new physics below the Planck scale, i.e. the trans-Planckian or ultraviolet physics, leaves footprints on the CMB which is being observed today. This possibility was first proposed in \cite{Brandenberger:1999sw} and after that, has been discussed in many literatures \cite{Yusofi:2014mta,Bhattacharya:2005wn,Mottola:1984ar,Allen:1985ux,Kundu:2013gha,Niemeyer:2000eh,Brandenberger:2000wr,Martin:2000xs,Tanaka:2000jw,Hui:2001ce,Niemeyer:2001qe,Kaloper:2002uj,Kaloper:2002cs,Burgess:2002ub,Brandenberger:2002hs,Elgaroy:2003gq,Greene:2004np,Greene:2005wk,Danielsson:2002kx,Danielsson:2002mb,Danielsson:2002qh,Danielsson:2005cc,Danielsson:2006gg,Goldstein:2002fc,Easther:2002xe,Chung:2003wn,Kaloper:2003nv,Burgess:2003hw,Alberghi:2003am,Martin:2003kp,Meerburg:2010rp,Kundu:2011sg,Groeneboom:2007rf,Broy:2016zik,Matsui:2017ldm}. From the viewpoint of the choice of vacuum, the influence of trans-Planckian physics on the CMB was first done by Danielsson in \cite{Danielsson:2002kx} where he used the mixed states as the initial vacuum instead of traditionally Bunch-Davies vacuum \cite{Bunch:1978yq}.\\
It is of interest to consider Mukhanov-Sasaki equation \cite{Mukhanov:1990me,Baumann:2009ds} that governs the evolution of the scalar fluctuations. The initial condition of this equation is obtained by referring to the quantum nature of fluctuations. Traditionally, the initial condition is imposed in an infinite past when all fluctuation modes have infinitely small scale and the spacetime resembles Minkowskian. This leads to a unique vacuum, called the Bunch-Davies vacuum with the corresponding coefficients of mode functions, $A_q=(\sqrt{\pi}/2)e^{i\nu\pi/2+i\pi/4}$ and $B_q=0$ (See Eq. (\ref{constantsI})). However since the duration of inflationary era is finite, this choice of vacuum is not so accurate. Furthermore, one can not follow a given mode to the infinitely small scales and the ultra short distances are restricted to the Planck's length.\\
Besides the Bunch-Davies vacuum, there is also another commonly used prescription for vacuum definition and this is usually called the adiabatic vacuum \cite{Birrell:1982ix,Parker:1969au,Parker:1974qw}. The adiabatic vacuum is based on the notion of particle and as we know this notion is usually ill-defined in curved space-time. In adiabatic vacuum prescription, the second order differential equation that governs the evolution of the mode functions, is solved by using the WKB approximation. But the main point is that in most cases either the Bunch-Davies vacuum or the adiabatic vacuum do not specify vacuum uniquely \cite{Chung:2003wn}.\\
The coherent state, which is the reminder of classical harmonic oscillator, is a special class of initial states for inflaton. Since we know almost nothing about the physics of pre-inflationary era, there is no precedence between the various choices of initial conditions for this era. And a priori any excited state such as coherent state can be as good as any vacuum state such as Bunch-Davies vacuum. Furthermore, the coherent state can be considered as zero-point quantum fluctuations around some classical oscillation. In \cite{Kundu:2011sg,Kundu:2013gha} it has been shown that the power spectrum and the bi-spectrum for scalar and tensor fluctuations resulted from a large class of coherent states, are similar to those with Bunch-Davies vacuum state.\\
In the case of Bunch-Davies vacuum, the operator which annihilates the vacuum state of the quantum inflaton, is $\hat{a}_{\vec{q}}$, i.e. $\hat{a}_{\vec{q}}|0\rangle=0$. Another interesting possibility is that rather than $\hat{a}_{\vec{q}}$, it is a linear combination of $\hat{a}_{\vec{q}}$ and $\hat{a}^{\dagger}_{\vec{q}}$ which annihilates the vacuum state. These states are called the $\alpha$-states and are related to the Bunch-Davies state by Bogoliubov transformations. The bi-spectrum calculations, when the initial state for inflation is chosen to be $\alpha$-states, lead to an obvious deviations in some consistency relations, but these deviations are too small to be observed in the near future \cite{Kundu:2013gha}.\\
If inflation comes about in the presence of a thermal distribution of inflatons, the resulting vacuum is called the thermal vacuum. The basic assumption in this kind of vacuum  is the existence of a pre-inflationary radiation era. It is shown that by this choice of vacuum, the curvature power spectrum receives a temperature dependent factor and also the number of e-foldings in this case is bigger than the number of e-foldings that is needed for solving the horizon problem \cite{Bhattacharya:2005wn}.\\
Another vacuum prescription which has recently been introduced, is the excited-de Sitter modes \cite{Yusofi:2014mta}. It is based on the asymptotic expansion of Hankel functions (the solutions of Mukhanov-Saski equation to the first order in slow-roll parameters) up to the higher order of $1/q\tau$, where $q$ and $\tau$ are wave number of the perturbation mode and conformal time, respectively. In this case and for de Sitter inflation, where the Hubble parameter $H$ is constant, the leading correction of curvature power spectrum is of order $(H/\Lambda)^2$, in which $\Lambda$ is some scale that trans-Planckian effects emerge. We shall explain this point more clearly hereafter.\\
In Danielsson's  proposal \cite{Danielsson:2002kx}, the inaccessibility of Minkowskian vacuum in infinite past is legitimized by introducing the physical cutoff $\Lambda$, and it is assumed that the evolution of the perturbation modes begins once $q=a(t_i)\Lambda$, where $t_i$ is an initial time when initial condition should be imposed. For de Sitter inflation, this initial condition reads $\tau(t_i)=-\Lambda/qH$. The physical interpretation of this result is that for modes that have larger wavelengths today, we need to go further back in time to impose the initial condition. Imposing the initial condition at a finite time $t_i$ leads to corrections in scalar power spectrum which is of order $H/\Lambda$ \cite{Danielsson:2002kx}. Several other papers have estimated the corrections to scalar and tensor power spectra. In \cite{Kempf:2001fa} the magnitude of corrections is predicted to be of order $(H/\Lambda)^2$. Using a different approach, the authors of \cite{Easther:2001fi,Easther:2001fz} estimated that the corrections are of order $H/\Lambda$. Based on a low energy effective field theory approach in \cite{Kaloper:2002uj,Kaloper:2002cs}, it has been shown that the effects can not be larger than $(H/\Lambda)^2$. Unfortunately, there is no a unanimous consensus about the magnitude of these corrections. In this paper we use the Danielsson's $\alpha$-vacua prescription \cite{Danielsson:2002kx} and investigate how the inflationary predictions get modified.\\
The organization of this paper is as follows: In section \ref{sec:Quantum Scalar Field in Expanding Universe}, we present the canonical quantization of the free scalar field in FLRW background. By using the Heisenberg picture in which the operators are time dependent, we give a short review of Bogoliubov transformations which describe the connection between the creation and annihilation operators at different times. With these transformations, one can obtain a criterion for obtaining vacuum at an arbitrary time. In section \ref{sec:Mukhanov-Sasaki Equation in Slow-Roll Approximation}, using the Mukhanov-Sasaki equation in slow-roll regime, we discuss two possible vacuum conditions in detail, the Bunch-Davies vacuum and the $\alpha$-vacua respectively. Then for any choice, both scalar and tensor fluctuations is studied in section \ref{sec:Vacuum Choices}. Then in \ref{subsec:alpha-Vacua}, which is the main part of our work, we calculate the corrections to $n_s$ and $n_t$ due to $\alpha$-vacua in the slow-roll approximation. In section \ref{sec:Comparison with Observational Data} we compare our results with observational data provided by Planck satellite. Finally we end up with conclusion in section \ref{sec:Conclusion}.
\section{Quantum Scalar Field in an Expanding Universe}
\label{sec:Quantum Scalar Field in Expanding Universe}
In this section we review preliminary results of the quantization of the scalar field in FLRW background and establish our notation. For more details see \cite{Polarski:1995jg,Danielsson:2002kx,Bergstrom:2002yd,Kiefer:1998jk,Lesgourgues:1996jc,Danielsson:2002qh,Mukhanov:2007zz}. We begin by introducing a real massless scalar field $\phi$ defined by the action
\begin{eqnarray}\label{actionI}
S=\frac{1}{2}\int d^4x\sqrt{-g}g^{\mu\nu}\partial_\mu\phi\partial_\nu\phi
\end{eqnarray}
and we choose our background to be flat FLRW
\begin{eqnarray}\label{FLRW}
ds^2=dt^2-a^2(t)d\vec{x}^2
\end{eqnarray}
Converting the cosmic time to the conformal time, $d\tau=dt/a(t)$ and defining the rescaled field $u(\tau,\vec{x})=a(\tau)\phi(\tau,\vec{x})$, we get
\begin{eqnarray}\label{actionII}
S=\frac{1}{2}\int d^3\vec{x}d\tau\left(u'^2-2\frac{a'}{a}uu'+\frac{a'^2}{a^2}u^2-\abs{\vec{\nabla}u}^2\right)
\end{eqnarray}
where a prime stands for derivation with respect to the conformal time. The canonical momentum conjugate to the field $u$ is
\begin{eqnarray}\label{canonicalmomentumI}
\pi=\frac{\partial\mathcal{L}}{\partial u'}=u'-\frac{a'}{a}u
\end{eqnarray}
Applying the Euler-Lagrange equation to (\ref{actionII}), gives the following equation of motion
\begin{eqnarray}\label{equationofmotionI}
u''-\frac{a''}{a}u-\vec{\nabla}^2u=0
\end{eqnarray}
Using the standard convention for the Fourier transform
\begin{eqnarray}\label{Fourierconvention}
u(\tau,\vec{x})=\int\frac{d^3\vec{q}}{(2\pi)^{3/2}}u_{\vec{q}}(\tau)e^{i\vec{q}.\vec{x}}
\end{eqnarray}
in Eqs. (\ref{canonicalmomentumI}) and (\ref{equationofmotionI}), we conclude that
\begin{eqnarray}\label{canonicalmomentumII}
\pi_{\vec{q}}(\tau)=u'_{\vec{q}}(\tau)-\frac{a'}{a}u_{\vec{q}}(\tau)
\end{eqnarray}
and also each Fourier mode satisfies
\begin{eqnarray}\label{equationofmotionII}
u''_{\vec{q}}+(q^2-\frac{a''}{a})u_{\vec{q}}=0
\end{eqnarray}
Using the above relations, one can easily compute the classical Hamiltonian 
\begin{eqnarray}\label{HamiltonianI}
\begin{split}
H=\frac{1}{2}\int d^3\vec{q}\Big[&\pi_{\vec{q}}(\tau)\pi^{\star}_{\vec{q}}(\tau)+q^2u_{\vec{q}}(\tau)u^{\star}_{\vec{q}}(\tau)+\\&\frac{a'}{a}\left(u_{\vec{q}}(\tau)\pi^{\star}_{\vec{q}}(\tau)+\pi_{\vec{q}}(\tau)u^{\star}_{\vec{q}}(\tau)\right)\Big]
\end{split}
\end{eqnarray}
The above expression derived from the fact that the scalar field is real and therefore $u^{\star}_{\vec{q}}(\tau)=u_{-\vec{q}}(\tau)$. Now we promote the scalar field and its canonical conjugate momentum to quantum operators and impose the following equal-time commutation relations between them
\begin{eqnarray}\label{ETCRI}
\begin{split}
&[\hat{u}(\tau,\vec{x}),\hat{\pi}(\tau,\vec{y})]=i\delta^{(3)}(\vec{x}-\vec{y})\\&
[\hat{u}(\tau,\vec{x}),\hat{u}(\tau,\vec{y})]=[\hat{\pi}(\tau,\vec{x}),\hat\pi(\tau,\vec{y})]=0
\end{split}
\end{eqnarray}
The commutation relations in Eq. (\ref{ETCRI}) imply the following commutation relations between Fourier modes
\begin{eqnarray}\label{ETCRII}
\begin{split}
&[\hat{u}_{\vec{q}}(\tau),\hat{\pi}_{\vec{k}}(\tau)]=i\delta^{(3)}(\vec{q}+\vec{k})\\&
[\hat{u}_{\vec{q}}(\tau),\hat{u}_{\vec{k}}(\tau)]=[\hat{\pi}_{\vec{q}}(\tau),\hat{\pi}_{\vec{k}}(\tau)]=0
\end{split}
\end{eqnarray}
Now we define the time-dependent creation and annihilation operators as usual
\begin{eqnarray}\label{creationandannihilation}
\begin{split}
&\hat{u}_{\vec{q}}(\tau)=\frac{1}{\sqrt{2q}}\left(\hat{a}_{\vec{q}}(\tau)+\hat{a}^{\dagger}_{-\vec{q}}(\tau)\right)\\&
\hat{\pi}_{\vec{q}}(\tau)=-i\sqrt{\frac{q}{2}}\left(\hat{a}_{\vec{q}}(\tau)-\hat{a}^{\dagger}_{-\vec{q}}(\tau)\right)
\end{split}
\end{eqnarray}
Using Eq. (\ref{creationandannihilation}), the Hamiltonian expression (\ref{HamiltonianI}) can be written as
\begin{eqnarray}\label{HamiltonianII}
\begin{split}
H=\frac{1}{2}\int d^3\vec{q}\Big[&q\left(\hat{a}_{\vec{q}}(\tau)\hat{a}^{\dagger}_{\vec{q}}(\tau)+\hat{a}^{\dagger}_{\vec{q}}(\tau)\hat{a}_{\vec{q}}(\tau)\right)+\\&i\frac{a'}{a}\left(\hat{a}^{\dagger}_{\vec{q}}(\tau)\hat{a}^{\dagger}_{-\vec{q}}(\tau)-\hat{a}_{\vec{q}}(\tau)\hat{a}_{-\vec{q}}(\tau)\right)\Big]
\end{split}
\end{eqnarray}
where the time-dependent creation and annihilation operators must satisfy the following equal-time commutation relations
\begin{eqnarray}\label{ETCRIII}
\begin{split}
&[\hat{a}_{\vec{q}}(\tau),\hat{a}^{\dagger}_{\vec{k}}(\tau)]=\delta^{(3)}(\vec{q}-\vec{k})\\&
[\hat{a}_{\vec{q}}(\tau),\hat{a}_{\vec{k}}(\tau)]=[\hat{a}^{\dagger}_{\vec{q}}(\tau),\hat{a}^{\dagger}_{\vec{k}}(\tau)]=0
\end{split}
\end{eqnarray}
The creation and annihilation operators at any arbitrary time can be expanded in terms of their values at some fixed time $\tau_i$. To see this, we obtain the time evolution of these operators in the Heisenberg picture. From Hamiltonian (\ref{HamiltonianII}), one can immediately find that
\begin{eqnarray}\label{Heisenberg}
\begin{split}
&\frac{d}{d\tau}\hat{a}_{\vec{q}}(\tau)=-i\big[\hat{a}_{\vec{q}}(\tau),H\big]=-iq\hat{a}_{\vec{q}}(\tau)+\frac{a'}{a}\hat{a}^{\dagger}_{-\vec{q}}(\tau)\\&
\frac{d}{d\tau}\hat{a}^{\dagger}_{-\vec{q}}(\tau)=-i\big[\hat{a}^{\dagger}_{-\vec{q}}(\tau),H\big]=iq\hat{a}^{\dagger}_{-\vec{q}}(\tau)+\frac{a'}{a}\hat{a}_{\vec{q}}(\tau)
\end{split}
\end{eqnarray}
These equations have the following general solutions 
\begin{eqnarray}\label{BogoliubovI}
\begin{split}
&\hat{a}_{\vec{q}}(\tau)=\alpha_q(\tau)\hat{a}_{\vec{q}}(\tau_i)+\beta_q(\tau)\hat{a}^{\dagger}_{-\vec{q}}(\tau_i)\\&
\hat{a}^{\dagger}_{-\vec{q}}(\tau)=\alpha^{\star}_q(\tau)\hat{a}^{\dagger}_{-\vec{q}}(\tau_i)+\beta^{\star}_q(\tau)\hat{a}_{\vec{q}}(\tau_i)
\end{split}
\end{eqnarray}
which present the Bogoliubov transformations that describe the mixing of creation and annihilation operators with time. In Eq. (\ref{BogoliubovI}), $\alpha_q(\tau)$ and $\beta_q(\tau)$ are Bogoliubov coefficients and the asterisk means complex conjugation. Recalling Eq. (\ref{ETCRIII}), we conclude that
\begin{eqnarray}\label{normalizationI}
\abs{\alpha_q(\tau)}^2-\abs{\beta_q(\tau)}^2=1
\end{eqnarray}
Now, we define the mode functions $f_q(\tau)$ and $g_q(\tau)$ corresponding to the field operator and its canonical momentum through the relations
\begin{eqnarray}\label{modesI}
\begin{split}
&\hat{u}_{\vec{q}}(\tau)=f_q(\tau)\hat{a}_{\vec{q}}(\tau_i)+f^{\star}_q(\tau)\hat{a}^{\dagger}_{-\vec{q}}(\tau_i)\\&
\hat{\pi}_{\vec{q}}(\tau)=-i\left(g_q(\tau)\hat{a}_{\vec{q}}(\tau_i)-g^{\star}_q(\tau)\hat{a}^{\dagger}_{-\vec{q}}(\tau_i)\right)
\end{split}
\end{eqnarray}
where $f_q(\tau)$ must satisfy (\ref{equationofmotionII}). Substituting Eq. (\ref{BogoliubovI}) into Eq. (\ref{creationandannihilation}) and noting Eq. (\ref{modesI}), we find that
\begin{eqnarray}\label{modesII}
\begin{split}
&f_q(\tau)=\frac{1}{\sqrt{2q}}\left(\alpha_q(\tau)+\beta^{\star}_q(\tau)\right)\\&
g_q(\tau)=\sqrt{\frac{q}{2}}\left(\alpha_q(\tau)-\beta^{\star}_q(\tau)\right)=i\left(f'_q(\tau)-\frac{a'}{a}f_q(\tau)\right)
\end{split}
\end{eqnarray}
Also the commutation relations in Eqs. (\ref{ETCRII}) and (\ref{ETCRIII}) yield the following Wronskian condition
\begin{eqnarray}\label{Wronskian}
g_q(\tau)f^{\star}_q(\tau)+g^{\star}_q(\tau)f_q(\tau)=i\left(f'_q(\tau)f^{\star}_q(\tau)-f'^{\star}_q(\tau)f_q(\tau)\right)=1
\end{eqnarray}
We are now in a position to introduce the vacuum state. To do this, a question may be raised here. Which are the alternative choices of vacuum state? The simplest and most natural choice is
\begin{eqnarray}\label{vacuumI}
\hat{a}_{\vec{q}}(\tau_i)|0,\tau_i\rangle=0
\end{eqnarray}
By using Eq. (\ref{BogoliubovI}), the above condition reads
\begin{eqnarray}\label{vacuumII}
\alpha_q(\tau_i)\hat{a}_{\vec{q}}(\tau_i)|0,\tau_i\rangle+\beta_q(\tau_i)\hat{a}^{\dagger}_{-\vec{q}}(\tau_i)|0,\tau_i\rangle=0
\end{eqnarray}
The first term on the left hand side (LHS) of Eq. (\ref{vacuumII}) is zero thanks to Eq. (\ref{vacuumI}). So, in order to have a consistent definition for the vacuum state, we must demand that $\beta_q(\tau_i)=0$. The situation in which $\tau_i\rightarrow-\infty$ is called the Bunch-Davies vacuum \cite{Bunch:1978yq}. This choice corresponds to the existence of an asymptotically Minkowskian past. Since the duration of inflationary era is finite, imposing the boundary condition at infinite past is illogical. So the modification of the Bunch-Davies vacuum is inevitable. Another and more reasonable possibility is that $\tau_i$ in Eq. (\ref{vacuumI}) be some finite value. This choice of vacuum is called the $\alpha$-vacua \cite{Allen:1985ux,Mottola:1984ar,Danielsson:2002mb,Handley:2016ods}. We will return to this point in section \ref{sec:Vacuum Choices}.
\section{Mukhanov-Sasaki Equation in Slow-Roll Approximation}
\label{sec:Mukhanov-Sasaki Equation in Slow-Roll Approximation}
In this section, we will work with the so-called Mukhanov-Sasaki equation that describes the time evolution of scalar fluctuations during the inflationary era \cite{Baumann:2009ds,Weinberg:2008zzc}
\begin{eqnarray}\label{Mukhanov-SasakiequationI}
u''^{(S)}_{\vec{q}}+(q^2-\frac{z''}{z})u^{(S)}_{\vec{q}}=0
\end{eqnarray}
where $z\equiv a\dot{\bar{\phi}}/H$, dot means derivative with respect to the cosmic time, $\bar{\phi}(x)$ is the background inflaton field, $u^{(S)}_{\vec{q}}$ is the gauge invariant perturbation variable  and superscript $S$ stands for the case of scalar fluctuations. Eq. (\ref{Mukhanov-SasakiequationI}) can be converted to an equivalent form by defining $\mathcal{R}_{\vec{q}}(\tau)=u^{(S)}_{\vec{q}}(\tau)/z(\tau)$
\begin{eqnarray}\label{Mukhanov-SasakiequationII}
\mathcal{R}''_{\vec{q}}+2\frac{z'}{z}\mathcal{R}_{\vec{q}}'+q^2\mathcal{R}_{\vec{q}}=0
\end{eqnarray}
where $\mathcal{R}$ is the comoving curvature perturbation. Working with Eq. (\ref{Mukhanov-SasakiequationI}) it is necessary to calculate the quantity $z''/z$. Let us begin by recalling the definition of the so-called slow-roll parameters
\begin{eqnarray}\label{slow-rollparametersI}
\epsilon\equiv-\frac{\dot H}{H^2}~~;~~\eta\equiv\frac{\ddot H}{2H\dot H}~~;~~\xi\equiv\frac{\dddot{H}}{H\ddot{H}}~~;~~\zeta\equiv\frac{\ddddot H}{H\dddot H}
\end{eqnarray}
and a useful combination of two first parameters defined as $\delta\equiv \eta-\epsilon$. In terms of the conformal time, the slow-roll parameters read
\begin{eqnarray}\label{slow-rollparametersII}
\epsilon=1-\frac{\mathcal{H}'}{\mathcal{H}^2}~~;~~\delta=1+\epsilon-\frac{z'}{\mathcal{H}z}
\end{eqnarray}
where $\mathcal{H}=aH$ and in the second expression we have used $z'/{\mathcal{H}z}=\epsilon+{\bar{\phi}''}/{\mathcal{H}\bar{\phi}'}$ and the definition of $z$ in terms of conformal time, $z=a\bar{\phi}'/\mathcal{H}$. From Eq. (\ref{slow-rollparametersI}),  the first  derivatives of slow-roll parameters are of the second order
\begin{eqnarray}\label{deriv}
\dot\epsilon=2H\epsilon(\epsilon+\eta)~~;~~\dot{\eta}=\eta(-2\eta+\epsilon+\xi)H~~;~~\dot\xi=\xi(\epsilon-\xi+\zeta)H
\end{eqnarray}
and can be written in the form
\begin{eqnarray}\label{derivativeofepsilon}
\epsilon'=2\mathcal{H}\epsilon(\epsilon+\eta)~~;~~\delta'=\epsilon'-\frac{z''}{\mathcal{H}z}+\frac{z'\mathcal{H}'}{\mathcal{H}^2z}+\frac{(z')^2}{\mathcal{H}z^2}
\end{eqnarray}
Thus we have
\begin{eqnarray}\label{computationIII}
\frac{z''}{\mathcal{H}z}=\frac{z'\mathcal{H}'}{\mathcal{H}^2z}+\frac{(z')^2}{\mathcal{H}z^2}+\mathcal{O}(\epsilon^2,\eta^2)
\end{eqnarray}
Substituting Eqs. (\ref{slow-rollparametersII}) into Eq. (\ref{computationIII}) and keeping only terms that are first order in slow-roll parameters, lead to
\begin{eqnarray}\label{computationIV}
\frac{z''}{z}=\mathcal{H}^2(2+5\epsilon-3\eta)
\end{eqnarray}
To express $\mathcal{H}$ in terms of slow-roll parameters, we integrate Eq. (\ref{slow-rollparametersII}) and choose suitable integration constants. This  yields
\begin{eqnarray}\label{aHsolution}
\mathcal{H}=-\frac{1}{(1-\epsilon)\tau}
\end{eqnarray}
Substituting these two latter equations into (\ref{Mukhanov-SasakiequationI}), we finally obtain the  Mukhanov-Sasaki equation to the first order in slow-roll parameters
\begin{eqnarray}\label{firstorderMukhanov-Sasakiequation}
u''^{(S)}_{\vec{q}}+(q^2-\frac{2+9\epsilon-3\eta}{\tau^2})u^{(S)}_{\vec{q}}=0
\end{eqnarray}
For constant $\epsilon$ and $\eta$, the general solution of Eq. (\ref{firstorderMukhanov-Sasakiequation}) is
\begin{eqnarray}\label{firstorderMukhanov-SasakisolutionI}
u^{(S)}_{\vec{q}}(\tau)=\sqrt{-\tau}[A_qH_\nu^{(1)}(-q\tau)+B_qH_\nu^{(2)}(-q\tau)]
\end{eqnarray}
where $H_\nu^{(1)}(-q\tau)$ and $H_\nu^{(2)}(-q\tau)$ are Hankel functions of first and second kind respectively. $A_q$ and $B_q$ are $q$-dependent constants of integration and $\nu\approx\frac{3}{2}+3\epsilon-\eta$ up to the first order in slow-roll parameters. Using the general solution (\ref{firstorderMukhanov-SasakisolutionI}) and the definition of $g^{(S)}_q(\tau)$ in Eq. (\ref{modesII}), we obtain
\begin{eqnarray}\label{modesIII}
\begin{split}
g^{(S)}_q(\tau)=&\frac{i}{\sqrt{-\tau}}\Big(A_qq\tau H^{(1)}_{\nu-1}(-q\tau)+A_q(2\epsilon-\eta)H^{(1)}_{\nu}(-q\tau)+\\&B_qq\tau H^{(2)}_{\nu-1}(-q\tau)+B_q(2\epsilon-\eta)H^{(2)}_{\nu}(-q\tau)\Big)
\end{split}
\end{eqnarray}
where we have ignored second and higher order terms in slow-roll parameters and the following recurrence relations between Hankel functions and their derivatives have been used \cite{Arfken:2012}
\begin{eqnarray}\label{Hankelrecurrence}
\begin{split}
&\frac{d}{dx}H^{(1,2)}_\nu(x)=\frac{1}{2}\left(H^{(1,2)}_{\nu-1}(x)-H^{(1,2)}_{\nu+1}(x)\right)\\&
H^{(1,2)}_{\nu-1}(x)+H^{(1,2)}_{\nu+1}(x)=\frac{2\nu}{x}H^{(1,2)}_\nu(x)
\end{split}
\end{eqnarray}
Now we can easily calculate the Bogoliubov coefficients from Eq. (\ref{modesII}). The result is
\begin{eqnarray}\label{BogoliubovII}
\begin{split}
\beta^{\star(S)}_q(\tau)=&\sqrt{\frac{q}{2}}f^{(S)}_q(\tau)-\frac{1}{\sqrt{2q}}g^{(S)}_q(\tau)=\\&-\frac{i}{\sqrt{-2q\tau}}\big[A_qq\tau H^{(1)}_{\nu-1}(-q\tau)+A_q(2\epsilon-\eta-iq\tau)H^{(1)}_{\nu}(-q\tau)\\&+B_qq\tau H^{(2)}_{\nu-1}(-q\tau)+B_q(2\epsilon-\eta-iq\tau)H^{(2)}_{\nu}(-q\tau)\big]
\end{split}
\end{eqnarray}
and for $\alpha^{(S)}_q(\tau)$ we get
\begin{eqnarray}\label{BogoliubovIII}
\begin{split}
\alpha^{(S)}_q(\tau)=&\sqrt{\frac{q}{2}}f^{(S)}_q(\tau)+\frac{1}{\sqrt{2q}}g^{(S)}_q(\tau)=\\&\frac{i}{\sqrt{-2q\tau}}\big[A_qq\tau H^{(1)}_{\nu-1}(-q\tau)+A_q(2\epsilon-\eta+iq\tau)H^{(1)}_{\nu}(-q\tau)\\&+B_qq\tau H^{(2)}_{\nu-1}(-q\tau)+B_q(2\epsilon-\eta+iq\tau)H^{(2)}_{\nu}(-q\tau)\big]
\end{split}
\end{eqnarray}
Similarly, the differential equation that governs tensor fluctuations during inflation is the same as the Mukhanov-Sasaki equation (\ref{Mukhanov-SasakiequationI}), except that $z$ variable is replaced with the scale factor \cite{Weinberg:2008zzc}. Therefore we have
\begin{eqnarray}\label{tensorfluctuations}
u''^{(T)}_{\vec{q}}+(q^2-\frac{a''}{a})u^{(T)}_{\vec{q}}=0
\end{eqnarray}
in which superscript $T$ stands for the tensorial fluctuations. Using $\mathcal{D}_{\vec{q}}(\tau)=u^{(T)}_{\vec{q}}(\tau)/a(\tau)$, the above equation can be written as
\begin{eqnarray}\label{Mukhanov-SasakiequII}
\mathcal{D}''_{\vec{q}}+2\mathcal{H}\mathcal{D}_{\vec{q}}'+q^2\mathcal{D}_{\vec{q}}=0
\end{eqnarray}
where $\mathcal{D}_{\vec{q}}$ is the gauge invariant tensor amplitude. Considering the slow-roll approximation and using Eqs. (\ref{slow-rollparametersII}) and (\ref{aHsolution}), Eq. (\ref{tensorfluctuations}) reduces to
\begin{eqnarray}\label{firstorderMukhanov-Sasakiequationtensor}
u''^{(T)}_{\vec{q}}+(q^2-\frac{2+3\epsilon}{\tau^2})u^{(T)}_{\vec{q}}=0
\end{eqnarray}
Once again, for constant $\epsilon$, the solutions of the above equation are the linear combination of $\sqrt{-\tau} H^{(1)}_\mu(-q\tau)$ and $\sqrt{-\tau} H^{(2)}_\mu(-q\tau)$ where $\mu\approx\frac{3}{2}+\epsilon$ to first order in $\epsilon$.
Just as in the case of scalar perturbations, one can perform a similar calculation  to get the Bogoliubov coefficients, corresponding to (\ref{BogoliubovII}) and (\ref{BogoliubovIII}), for tensorial fluctuations. The result is
\begin{eqnarray}\label{BogoliubovIV}
\begin{split}
\beta^{\star(T)}_q(\tau)=\sqrt{\frac{q\tau}{2}}\Big[&C_q\left(H^{(1)}_{\mu-1}(-q\tau)-iH^{(1)}_{\mu}(-q\tau)\right)+\\&D_q\left(H^{(2)}_{\mu-1}(-q\tau)-iH^{(2)}_{\mu}(-q\tau)\right)\Big]\\
\alpha^{(T)}_q(\tau)=-\sqrt{\frac{q\tau}{2}}\Big[&C_q\left(H^{(1)}_{\mu-1}(-q\tau)+iH^{(1)}_{\mu}(-q\tau)\right)+\\&D_q\left(H^{(2)}_{\mu-1}(-q\tau)+iH^{(2)}_{\mu}(-q\tau)\right)\Big]
\end{split}
\end{eqnarray}
in which $C_q$ and $D_q$ are $q$-dependent constants.
\section{Vacuum Choices}
\label{sec:Vacuum Choices}
In order to define the vacuum state, we need to fix the mode functions. Here, we will consider two distinct vacuum states and calculate scalar and tensor power spectra in both cases.
\subsection{The Bunch-Davies Vacuum}
\label{sec:The Bunch-Davies Vacuum}
\paragraph{Scalar Fluctuations}
First, we consider the situation where the space-time resembles Minkowskian at very early times, called the Bunch-Davies vacuum \cite{Bunch:1978yq}. This requires that at sufficiently early times (large negative conformal time $\tau$), the mode function $f^{(S)}_q(\tau)$ behaves as \cite{Baumann:2009ds,Mukhanov:2007zz,Birrell:1982ix,Weinberg:2008zzc}
\begin{eqnarray}\label{Bunch-Daviesvacuum}
\lim_{-\tau\to\infty}f^{(S)}_q(\tau)=\frac{1}{\sqrt{2q}}e^{-iq\tau}
\end{eqnarray}
Recalling the asymptotic behavior of the Hankel functions for large real argument
\begin{eqnarray}\label{largeasymptotic}
H_\nu^{(1)}(x)\longrightarrow\sqrt{\frac{2}{\pi x}}\exp(ix-i\nu\frac{\pi}{2}-i\frac{\pi}{4})~~~;~~~H_\nu^{(2)}(x)=H_\nu^{(1)\star}(x)
\end{eqnarray}
and comparing Eq. (\ref{firstorderMukhanov-SasakisolutionI}) with Eq. (\ref{Bunch-Daviesvacuum}), we infer that
\begin{eqnarray}\label{constantsI}
A_q=\frac{\sqrt{\pi}}{2}e^{i\nu\pi/2+i\pi/4}~~~and~~~B_q=0
\end{eqnarray}
So Eq. (\ref{firstorderMukhanov-SasakisolutionI}) reduces to
\begin{eqnarray}\label{firstorderMukhanov-SasakisolutionII}
f^{(S)}_q(\tau)=\frac{\sqrt{-\pi\tau}}{2}e^{i\nu\pi/2+i\pi/4}H^{(1)}_\nu(-q\tau)
\end{eqnarray}
To check the consistency of the above result with the procedure that has been explained in section \ref{sec:Quantum Scalar Field in Expanding Universe}, we use Eqs. (\ref{BogoliubovII}), (\ref{BogoliubovIII}), and (\ref{constantsI}) and calculate $\beta^{\star(S)}_q(\tau)$ and $\alpha^{(S)}_q(\tau)$. The result for $\beta^{\star(S)}_q(\tau)$ is
\begin{eqnarray}\label{consistencyI}
\begin{split}
\beta^{\star(S)}_q(\tau)=&-\frac{1}{4}\sqrt{\frac{\pi}{-2q\tau}}\big[(-2iq\tau+3\pi\epsilon q\tau-\pi\eta q\tau)H^{(1)}_{\nu-1}(-q\tau)+\\&(-4i\epsilon+2i\eta-2q\tau-3i\pi\epsilon q\tau+i\pi\eta q \tau)H^{(1)}_\nu(-q\tau)\big]
\end{split}
\end{eqnarray}
and for $\alpha^{(S)}_q(\tau)$
\begin{eqnarray}\label{consistencyII}
\begin{split}
\alpha^{(S)}_q(\tau)=&\frac{1}{4}\sqrt{\frac{\pi}{-2q\tau}}\big[(-2iq\tau+3\pi\epsilon q\tau-\pi\eta q\tau)H^{(1)}_{\nu-1}(-q\tau)+\\&(-4i\epsilon+2i\eta+2q\tau+3i\pi\epsilon q\tau-i\pi\eta q \tau)H^{(1)}_\nu(-q\tau)\big]
\end{split}
\end{eqnarray}
It is easy to check that the right hand side of Eq. (\ref{consistencyI}) goes to zero as $\tau\rightarrow-\infty$. So we have $\beta^{(S)}_q(\tau_i\rightarrow-\infty)=0$, as it should be. For small real argument, the first Hankel function behaves as: $H^{(1)}_\nu(x)\longrightarrow{-i\Gamma(\nu)}({x}/{2})^{-\nu}/{\pi}$.
So, beyond the horizon, when $q/aH\ll1$, Eq. (\ref{firstorderMukhanov-SasakisolutionII}) leads to the following asymptotic form
\begin{eqnarray}\label{firstorderMukhanov-SasakisolutionIII}
\left.f^{(S)}_q(\tau)\right\vert_{q/aH\ll1}=\frac{1-i}{\sqrt{\pi}}e^{i\nu\pi/2}2^{\nu-3/2}\Gamma(\nu)q^{-\nu}(-\tau)^{1/2-\nu}
\end{eqnarray}
Now let us to consider the power spectrum of scalar fluctuations defined as
\begin{eqnarray}\label{powerspectrumI}
\mathcal{P}^{(S)}_{\mathcal{R}}(\tau,q)\equiv \frac{q^3}{2\pi^2}\abs{\mathcal{R}_q(\tau)}^2=\frac{q^3}{2\pi^2}\frac{1}{z^2}\abs{f^{(S)}_q(\tau)}^2
\end{eqnarray}
and gives a gauge invariant measure of fluctuations. Thus from Eqs. (\ref{firstorderMukhanov-SasakisolutionIII}) and (\ref{powerspectrumI}) we obtain
\begin{eqnarray}\label{powerspectrumIII}
\mathcal{P}^{(S)}_{\mathcal{R},0}(\tau,q)=\frac{1}{z^2}\frac{1}{\pi^3}2^{2\nu-3}[\Gamma(\nu)]^2q^{-2\nu+3}(-\tau)^{1-2\nu}
\end{eqnarray}
where subscript $0$ denotes that the power spectrum is evaluated outside the horizon. Using Eqs. (\ref{slow-rollparametersII}) and (\ref{aHsolution}) we get
\begin{eqnarray}\label{zequation}
\frac{z'}{z}=-\frac{1}{(1-\epsilon)\tau}(1+2\epsilon-\eta)
\end{eqnarray}
and thus $z\propto(-\tau)^{1/2-\nu}$. Substituting this in Eq. (\ref{powerspectrumIII}), we conclude that, to first order in slow-roll parameters, the curvature perturbation and the corresponding power spectrum remain nearly constant at superhorizon scales. This means that one can calculate it at any convenient time which is usually chosen the time of horizon crossing, $t_q$,  where $q/a(t_q)=H(t_q)$ or $q=\mathcal{H}(t_q)$. To zeroth order in slow-roll parameters Eq. (\ref{aHsolution}) leads to
\begin{eqnarray}\label{tauatcrossing}
\tau(t_q)=-\frac{1}{\big(1-\epsilon(t_q)\big)q}\simeq-\frac{1}{q}
\end{eqnarray}
Also since $\dot H=-4\pi G\dot{\bar\phi}^2$, Eq. (\ref{slow-rollparametersI}) yields
\begin{eqnarray}\label{phiatcrossing}
\dot{\bar\phi}(t_q)=\pm\sqrt{\frac{-\dot H(t_q)}{4\pi G}}=\pm H(t_q)\sqrt{\frac{\epsilon(t_q)}{4\pi G}}
\end{eqnarray}
So we have
\begin{eqnarray}\label{zatcrossing}
z(t_q)=\pm\frac{q}{H(t_q)}\sqrt{\frac{\epsilon(t_q)}{4\pi G}}
\end{eqnarray}
Putting Eqs. (\ref{tauatcrossing}) and (\ref{zatcrossing}) into Eq. (\ref{powerspectrumIII}) we obtain
\begin{eqnarray}\label{powerspectrumatcrossing}
\mathcal{P}^{(S)}_{\mathcal{R},0}(t_q,q=\mathcal{H}(t_q))=\frac{H^2(t_q)}{\epsilon(t_q)}\frac{G}{\pi^2}2^{2\nu-1}[\Gamma(\nu)]^2
\end{eqnarray}
In order to find the $q$-dependence of above expression, the q-dependence of $H(t_q)$ and $\epsilon(t_q)$ must be specified. To find these, we begin by differentiating the horizon crossing condition, $q/a(t_q)=H(t_q)$, with respect to $q$ \cite{Weinberg:2008zzc}
\begin{eqnarray}\label{q-dependenceI}
\frac{dt_q}{dq}=\frac{1}{a(t_q)[H^2(t_q)+\dot H(t_q)]}
\end{eqnarray}
Then, from definition of the first slow-roll parameter in Eq. (\ref{slow-rollparametersI}), we get
\begin{eqnarray}\label{q-dependenceII}
\frac{q}{H(t_q)}\frac{dH(t_q)}{dq}=-\frac{\epsilon(t_q)}{1-\epsilon(t_q)}
\end{eqnarray}
Assuming that $\epsilon(t_q)$ is small, we have
\begin{eqnarray}\label{footnoteI}
\frac{dH(t_q)}{H(t_q)}\approx-\epsilon(t_q)\frac{dq}{q}
\end{eqnarray}
Integrating the above equation and ignoring the $q$-dependency of $\epsilon(t_q)$, lead to 
\begin{eqnarray}\label{q-dependenceIII}
H(t_q)=H_{l}(\frac{q}{q_{l}})^{-\epsilon}
\end{eqnarray}
In the above equation $H_{l}$ is the value of Hubble parameter at the time when the last scale, $q_{l}$, leaves the horizon. According to this equation, the fluctuations with smaller wavelengths leave the horizon later. Now, in the following, we calculate the $q$-dependence of $\epsilon(t_q)$. To do this, we use Eqs. (\ref{deriv}) and (\ref{q-dependenceI}) which give
\begin{eqnarray}\label{q-dependenceIV}
\frac{q}{\epsilon(t_q)}\frac{d\epsilon(t_q)}{dq}=\frac{2[\epsilon(t_q)+\eta(t_q)]}{1-\epsilon(t_q)}
\end{eqnarray}
so
\begin{eqnarray}\label{q-dependenceV}
\epsilon(t_q)=\epsilon_{l}(\frac{q}{q_{l}})^{2\epsilon+2\eta}
\end{eqnarray}
where $\epsilon_{l}$ is the value of the first slow-roll parameter when the last scale leaves the horizon, i.e. $\epsilon_{l}\equiv\epsilon(t_q(q_{l}))$. If we assume that the last scale has left the horizon, almost at the end of inflation, then $\epsilon_{l}$ is typically a number of order unity. For completeness sake, we derive the $q$-dependence of $\eta(t_q)$ and $\xi(t_q)$ in a similar way. In this case Eqs. (\ref{deriv}) and (\ref{q-dependenceI}) lead to
\begin{eqnarray}\label{etaatcrossingdependenceI}
\begin{split}
&\frac{d\eta(t_q)}{dq}=\eta(t_q)[\epsilon(t_q)-2\eta(t_q)+\xi(t_q)]\frac{1}{q}\\&
\frac{d\xi(t_q)}{dq}=\xi(t_q)[\epsilon(t_q)-\xi(t_q)+\zeta(t_q)]\frac{1}{q}
\end{split}
\end{eqnarray}
which after integrating give
\begin{eqnarray}\label{etaatcrossingdependenceII}
\begin{split}
&\eta(t_q)=\eta_{l}(\frac{q}{q_{l}})^{\epsilon-2\eta+\xi}\\&
\xi(t_q)=\xi_{l}(\frac{q}{q_{l}})^{\epsilon-\xi+\zeta}
\end{split}
\end{eqnarray}
and as usual, $\eta_{l}$ and $\xi_{l}$ are the values of $\eta (t)$ and $\xi(t)$ when the last scale leaves the horizon. By substituting Eqs. (\ref{q-dependenceIII}) and (\ref{q-dependenceV}) into (\ref{powerspectrumatcrossing}), we obtain the power spectrum of scalar fluctuations
\begin{eqnarray}\label{powerspectrumDr.}
\mathcal{P}^{(S)}_{\mathcal{R},0}(t_q,q=\mathcal{H}(t_q))=\frac{G}{\pi^2}\frac{H^2_{l}}{\epsilon_{l}}2^{2\nu-1}[\Gamma(\nu)]^2(\frac{q}{q_{l}})^{-4\epsilon-2\eta}
\end{eqnarray}
Now using the definition of the spectral index $n_s$
\begin{eqnarray}\label{scalarspectralindexI}
n_s(q)-1\equiv\left(\frac{q}{\mathcal{P}^{(S)}_{\mathcal{R},0}(t_q,q)}\frac{d\mathcal{P}^{(S)}_{\mathcal{R},0}(t_q,q)}{dq}\right)_{q={\cal H}(t_q)}
\end{eqnarray}
we can easily obtain
\begin{eqnarray}\label{scalarspectralindexII}
n_s(q)=1-4\epsilon(t_q)-2\eta(t_q)
\end{eqnarray}
Now Eqs. (\ref{q-dependenceV}) and (\ref{etaatcrossingdependenceII}) can be used to find the $q$-dependence of the scalar spectral index as
\begin{eqnarray}\label{scalarspectralindexIII}
n_s(q)=1-4\epsilon_{l}(\frac{q}{q_{l}})^{2\epsilon+2\eta}-2\eta_{l}(\frac{q}{q_{l}})^{\epsilon-2\eta+\xi}
\end{eqnarray}
\paragraph{Tensor Fluctuations}
Imposing the Bunch-Davies initial condition on the mode functions for tensor fluctuations that is given by the solution of Eq. (\ref{firstorderMukhanov-Sasakiequationtensor}), yields \cite{Baumann:2009ds,Mukhanov:2007zz,Birrell:1982ix,Weinberg:2008zzc}
\begin{eqnarray}\label{tensorsolutionI}
f^{(T)}_q(\tau)=\frac{\sqrt{-\pi\tau}}{2}H^{(1)}_\mu(-q\tau)
\end{eqnarray}
And the power spectrum of tensor fluctuation is defined as
\begin{eqnarray}\label{tensorpowerspectrumI}
\mathcal{P}^{(T)}_{\mathcal{D}}(\tau,q)\equiv \frac{32Gq^3}{\pi a^2}\abs{f^{(T)}_q(\tau)}^2
\end{eqnarray}
As before, by using Eq. (\ref{q-dependenceIII}), one can evaluate this expression outside the horizon and then it is worthwhile to rewrite the conserved power spectrum in terms of quantities at the time of horizon crossing
\begin{eqnarray}\label{tensorpowerspectrumII}
\mathcal{P}^{(T)}_{\mathcal{D},0}(t_q,q=\mathcal H(t_q))=\frac{32GH^2_{l}}{\pi^2}2^{2\mu-2}[\Gamma(\mu)]^2(\frac{q}{q_{l}})^{-2\epsilon}
\end{eqnarray}
The tensor spectral index is defined as
\begin{eqnarray}\label{tensorspectralindexI}
n_t(q)\equiv\left(\frac{d\ln \mathcal{P}^{(T)}_{\mathcal{D},0}(t_q,q)}{d\ln q}\right)_{q={\cal H}(t_q)}
\end{eqnarray}
So by using Eq. (\ref{q-dependenceV}) we obtain
\begin{eqnarray}\label{tensorspectralindexII}
n_t(q)=-2\epsilon(t_q)=-2\epsilon_{l}(\frac{q}{q_{l}})^{-2\epsilon}
\end{eqnarray}
\subsection{$\alpha$-Vacua}
\label{subsec:alpha-Vacua}
As stated before, since the duration of inflation is finite, imposing the Minkowskian vacuum at infinite past is not legitimate. So, the initial condition should be imposed at some finite time $\tau_i$ \cite{Danielsson:2002kx}. This is called $\alpha$-vacua and thus Bunch-Davies vacuum is a special case of it when $\tau_i\rightarrow-\infty$.
\paragraph{Scalar Fluctuations}
Recalling the definition of vacuum state, Eq. (\ref{vacuumI}), we demand that at some arbitrary time, say $\tau_i$, $\beta_q(\tau_i)$ must be zero. Imposing this condition on Eq. (\ref{BogoliubovII}), we obtain
\begin{eqnarray}\label{vacuumconditionI}
\begin{split}
\beta^{\star}_q(\tau_i)=&-\frac{i}{\sqrt{-2q\tau_i}}\\&\big[A_qq\tau_iH^{(1)}_{\nu-1}(-q\tau_i)+A_q\left(2\epsilon(\tau_i)-\eta(\tau_i)-iq\tau_i\right)H^{(1)}_{\nu}(-q\tau_i)+\\&~B_qq\tau_iH^{(2)}_{\nu-1}(-q\tau_i)+B_q\left(2\epsilon(\tau_i)-\eta(\tau_i)-iq\tau_i\right)H^{(2)}_{\nu}(-q\tau_i)\big]=0
\end{split}
\end{eqnarray}
Solving this gives
\begin{eqnarray}\label{vacuumconditionII}
B_q=\frac{A_qe^{-2iq\tau_i}\left(2\epsilon(\tau_i)-\eta(\tau_i)\right)}{2iq\tau_i}
\end{eqnarray}
where we have assumed that $\abs{\tau_i}$ is large enough to use the asymptotic behavior of the Hankel functions with large argument. The special case of this relation was derived recently by \cite{Broy:2016zik} for  de Sitter inflation (for more detail see Eq. (39) of \cite{Broy:2016zik}). Using the normalization condition (\ref{normalizationI}), we get
\begin{eqnarray}\label{normalizationII}
\abs{A_q}^2=\frac{\pi}{4}
\end{eqnarray}
Then at sufficiently late times in the slow-roll era, we find that
\begin{eqnarray}\label{scalaralphavacuaI}
\left.f_q(\tau)\right\vert_{q/aH\ll1}=\frac{-i\Gamma(\nu)}{\pi}2^\nu q^{-\nu}(-\tau)^{1/2-\nu}(A_q-B_q)
\end{eqnarray}
One can easily check that $\mathcal{R}_q(\tau)$ is constant. As before, we calculate the power spectrum of scalar fluctuations at horizon crossing
\begin{eqnarray}\label{scalaralphavacuaII}
\begin{split}
^{\alpha}\mathcal{P}^{(S)}_{\mathcal{R},0}(t_q,q=\mathcal H(t_q))=&\frac{G}{\pi^3}\frac{H^2(t_q)}{\epsilon(t_q)}2^{2\nu+1}[\Gamma(\nu)]^2\times\\&\left(\abs{A_q}^2+\abs{B_q}^2-2\Re(A_qB^{\star}_q)\right)
\end{split}
\end{eqnarray}
where the left superscript $\alpha$ means that this is the power spectrum when the initial condition is chosen to be the $\alpha$-vacua. Substituting from Eqs. (\ref{vacuumconditionII}) and (\ref{normalizationII}), up to leading order in slow-roll parameters, one gets
\begin{eqnarray}\label{scalaralphavacuaIII}
\begin{split}
^{\alpha}\mathcal{P}^{(S)}_{\mathcal{R},0}(t_q,q=\mathcal H(t_q))=&\frac{G}{\pi^2}\frac{H^2_{l}}{\epsilon_{l}}2^{2\nu-1}[\Gamma(\nu)]^2(\frac{q}{q_{l}})^{-4\epsilon-2\eta}\times\\&\big[1+\frac{\left(2\epsilon(\tau_i)-\eta(\tau_i)\right)}{q\tau_i}\sin(2q\tau_i)\big]
\end{split}
\end{eqnarray}
Here it is worthwhile to notice that in above expression, $\tau_i$ is $q$-dependent (this dependence will be discussed shortly, see Eq. (\ref{Hubbleatfirstscale})). It is useful to write Eq. (\ref{scalaralphavacuaIII}) as standard power spectrum, given by Eq. (\ref{powerspectrumDr.}), plus an extra term that includes correction due to $\alpha$-vacua. So writing
\begin{eqnarray}\label{BroynotationI}
^{\alpha}\mathcal{P}^{(S)}_{\mathcal{R},0}(t_q,q)=\mathcal{P}^{(S)}_{\mathcal{R},0}(t_q,q)+\delta\mathcal{P}^{(S)}_{\mathcal{R},0}(t_q,q)
\end{eqnarray}
After substituting the above expression into Eq. (\ref{scalarspectralindexI}), we get
\begin{eqnarray}\label{BroynotationIV}
\begin{split}
n_s(q)-1=&\left(\frac{q}{\mathcal{P}^{(S)}_{\mathcal{R},0}(t_q,q)}\frac{d\mathcal{P}^{(S)}_{\mathcal{R},0}(t_q,q)}{dq}+\frac{q}{\mathcal{P}^{(S)}_{\mathcal{R},0}(t_q,q)}\frac{d\delta\mathcal{P}^{(S)}_{\mathcal{R},0}(t_q,q)}{dq}\right.\\&\left.-\frac{q}{\mathcal{P}^{(S)}_{\mathcal{R},0}(t_q,q)}\frac{\delta\mathcal{P}^{(S)}_{\mathcal{R},0}(t_q,q)}{\mathcal{P}^{(S)}_{\mathcal{R},0}(t_q,q)}\frac{d\mathcal{P}^{(S)}_{\mathcal{R},0}(t_q,q)}{dq}\right)_{q={\cal H}(t_q)}
\end{split}
\end{eqnarray}
to first order in correction terms and
\begin{eqnarray}\label{BroynotationII}
\begin{split}
\delta\mathcal{P}^{(S)}_{\mathcal{R},0}(t_q,q)=&\frac{G}{\pi^2}\frac{H^2_{l}}{\epsilon_{l}}2^{2\nu-1}[\Gamma(\nu)]^2(\frac{q}{q_{l}})^{-4\epsilon-2\eta}\times\\&\frac{\left(2\epsilon(\tau_i)-\eta(\tau_i)\right)}{q\tau_i}\sin(2q\tau_i)
\end{split}
\end{eqnarray}
As discussed before, in the case of $\alpha$-vacua it is impossible to access the Minkowskian vacuum at infinite past. We can rephrase this impossibility by introducing a physical cutoff which is described by some fixed scale of momentum, $\Lambda$, and assume that the mode evolution begins when $q=a(t_i)\Lambda$. Using this condition in Eq. (\ref{aHsolution}) we obtain
\begin{eqnarray}\label{alphavacuaconditionI}
\tau_i\equiv\tau(t_i)=-\frac{\Lambda}{[1-\epsilon(t_i)]qH(t_i)}
\end{eqnarray}
This means that we need to find the $q$-dependency of $H(t_i)$. Differentiating the initial condition, i.e. $q=a(t_i)\Lambda$, with respect to $q$ yields
\begin{eqnarray}\label{initialtimeI}
\frac{dt_i}{dq}=\frac{1}{\Lambda a(t_i)H(t_i)}
\end{eqnarray}
To find $H(t_i)$, we note that $\epsilon(t_i)$ is small and therefore Eq.(\ref{footnoteI}) is also satisfied for $t_i$. Thus
\begin{eqnarray}\label{Hubbleatfirstscale}
H(t_i)=H_{f}(\frac{q}{q_{f}})^{-\epsilon}
\end{eqnarray}
where $H_{f}$ is the value of Hubble parameter at the time when the first scale, $q_{f}$, satisfies the initial condition. According to Eq. (\ref{Hubbleatfirstscale}), wee see that the initial condition for the smaller wavelengths is imposed at later times. Comparing  Eqs. (\ref{q-dependenceIII}) and (\ref{Hubbleatfirstscale}) gives $H_{f}=H_{l}(q_{f}/q_{l})^{-\epsilon}$. Now substituting Eq. (\ref{Hubbleatfirstscale}) into Eq. (\ref{alphavacuaconditionI}), we see that to zeroth order in $\epsilon$
\begin{eqnarray}\label{alphavacuaconditionII}
\tau_i=-\frac{\Lambda q^{-\epsilon}_{f}}{H_{f}}q^{\epsilon-1}
\end{eqnarray}
The $q$-dependency of $\epsilon(t_i)$ and $\eta(t_i)$ can be calculated in a straightforward manner similar to what is done in Eqs.(\ref{q-dependenceIV}) and (\ref{q-dependenceV}) for $\epsilon(t_q)$. This yields
\begin{eqnarray}\label{initialtimeIV}
\epsilon(t_i)=\epsilon_{f}(\frac{q}{q_{f}})^{2\epsilon+2\eta}
\end{eqnarray}
where $\epsilon_{f}$ is the value of $\epsilon(t)$ when the first scale satisfies the initial condition and thus $\epsilon_{f}=\epsilon_{l}(q_{f}/q_{l})^{2\epsilon+2\eta}$. Performing the same steps that lead to Eqs. (\ref{etaatcrossingdependenceI}) and (\ref{etaatcrossingdependenceII}), it is easy to show that
\begin{eqnarray}\label{initialtimeVII}
\begin{split}
&\eta(t_i)=\eta_{f}(\frac{q}{q_{f}})^{\epsilon-2\eta+\xi}\\&
\xi(t_i)=\xi_{f}(\frac{q}{q_{f}})^{\epsilon-\xi+\zeta}
\end{split}
\end{eqnarray}
Eqs. (\ref{alphavacuaconditionII}), (\ref{initialtimeIV}), and the above equations, enable us to determine the $q$ dependence of the power spectrum (\ref{scalaralphavacuaIII}). We see that
\begin{eqnarray}\label{finalscalar}
\begin{split}
^{\alpha}\mathcal{P}^{(S)}_{\mathcal{R},0}(t_q,q=\mathcal H(t_q))=&\frac{G}{\pi^2}\frac{H^2_{l}}{\epsilon_{l}}2^{2\nu-1}[\Gamma(\nu)]^2(\frac{q}{q_{l}})^{-4\epsilon-2\eta}\times\\&\Big\{1+\frac{H_{f}}{\Lambda}\big[2\epsilon_{f}(\frac{q}{q_{f}})^{\epsilon+2\eta}-\eta_{f}(\frac{q}{q_{f}})^{-2\eta+\xi}\big]\times\\&\sin(\frac{2\Lambda(q/q_{f})^{\epsilon}}{H_{f}})\Big\}
\end{split}
\end{eqnarray}
The first term on the right hand side gives the power spectrum (\ref{powerspectrumDr.}) based on the Bunch-Davies vacuum as the initial state of the universe. The second term shows that correction to the power spectrum have a sinusoidal shape whose amplitude and frequency are scale dependent.\\
Now, it is straightforward to evaluate the scalar spectral index $n_s(q)$ which is resulted from Eq. (\ref{finalscalar}). Looking back to Eq. (\ref{scalarspectralindexIII}), since $(q/q_{l})^{2\epsilon+2\eta}$ and $(q/q_{l})^{\epsilon-2\eta+\xi}$ are very close to unity, the second and third term on the right hand side of Eq. (\ref{scalarspectralindexIII}) are of order $\epsilon_{l}$ and $\eta_{l}$. On the other hand, using (\ref{finalscalar}), the second term on the right hand side of Eq. (\ref{BroynotationIV}) is
\begin{eqnarray}\label{BroynotationV}
\begin{split}
&\frac{q}{\mathcal{P}^{(S)}_{\mathcal{R},0}(t_q,q)}\frac{d\delta\mathcal{P}^{(S)}_{\mathcal{R}}(t_q,q)}{dq}=2\epsilon_{f}\big[2\epsilon_{f}(\frac{q}{q_{f}})^{4\epsilon+4\eta}-\eta_{f}(\frac{q}{q_{f}})^{3\epsilon+\xi}\big]\times\\&\cos(\frac{2\Lambda(q/q_{f})^{\epsilon}}{H_{f}})+\frac{H_{f}}{\Lambda}\big[-6\epsilon^2_{f}(\frac{q}{q_{f}})^{3\epsilon+4\eta}+4\epsilon_{f}\eta_{f}(\frac{q}{q_{f}})^{2\epsilon+\xi}+\\&4\eta^2_{f}(\frac{q}{q_{f}})^{\epsilon-4\eta+2\xi}-\eta_{f}\xi_{f}(\frac{q}{q_{f}})^{\epsilon-2\eta+\zeta}\big]\sin(\frac{2\Lambda(q/q_{f})^{\epsilon}}{H_{f}})
\end{split}
\end{eqnarray}
The coefficient of sine in the above expression is second order in slow-roll parameters times a factor of $H_{f}/\Lambda$. Assuming that $\Lambda\gg H_{f}$ \cite{Danielsson:2002kx}, we can ignore this term in comparison with the first term. The third term in Eq. (\ref{BroynotationIV}) has a contribution with similar order of magnitude and we neglect it too. Collecting our results, we conclude that the scalar spectral index, when the initial condition is chosen to be $\alpha$-vacua, is
\begin{eqnarray}\label{spectralindexI}
\begin{split}
n_s(q)-1=&-4\epsilon_{l}(\frac{q}{q_{l}})^{2\epsilon+2\eta}-2\eta_{l}(\frac{q}{q_{l}})^{\epsilon-2\eta+\xi}+\\&2\epsilon_{f}\big[2\epsilon_{f}(\frac{q}{q_{f}})^{4\epsilon+4\eta}-\eta_{f}(\frac{q}{q_{f}})^{3\epsilon+\xi}\big]\cos(\frac{2\Lambda(q/q_{f})^{\epsilon}}{H_{f}})
\end{split}
\end{eqnarray}
\paragraph{Tensor Fluctuations}
In the case of tensor fluctuations, as it is mentioned before, the general solution of Eq. (\ref{firstorderMukhanov-Sasakiequationtensor}) is
\begin{eqnarray}\label{tensorII}
u^{(T)}_{\vec{q}}(\tau)=\sqrt{-\tau}[C_qH_\mu^{(1)}(-q\tau)+D_qH_\mu^{(2)}(-q\tau)]
\end{eqnarray}
Using the Bogoliubov coefficients of Eq. (\ref{BogoliubovIV}) and the $\alpha$-vacuum condition, $\beta^{\star}_q(\tau_i)=0$, lead to the following relation between $C_q$ and $D_q$
\begin{eqnarray}\label{vacuumconditionIII}
D_q=\frac{C_qe^{-2iq\tau_i}\left(2\epsilon(\tau_i)-\eta(\tau_i)\right)}{2iq\tau_i}
\end{eqnarray}
and the normalization condition in Eq. (\ref{normalizationI}) reads
\begin{eqnarray}\label{normalizationIII}
\abs{C_q}^2=\frac{\pi}{4}
\end{eqnarray}
Using Eq. (\ref{tensorpowerspectrumI}) and the asymptotic form of the Hankel functions for small real argument we calculate the power spectrum of tensor fluctuations outside the horizon
\begin{eqnarray}\label{tensoralphavacuaI}
^{\alpha}\mathcal{P}^{(T)}_{\mathcal{D},0}(\tau,q)=\frac{G}{\pi^3a^2}[\Gamma(\mu)]^22^{2\mu+5}q^{3-2\mu}(-\tau)^{1-2\mu}\abs{C_q-D_q}^2
\end{eqnarray}
As usual, by using Eq. (\ref{tauatcrossing}) and the horizon crossing condition,  we can evaluate the conserved power spectrum of tensor fluctuations at the horizon crossing
\begin{eqnarray}\label{tensoralphavacuaII}
\begin{split}
^{\alpha}\mathcal{P}^{(T)}_{\mathcal{D},0}(t_q,q=\mathcal{H}(t_q))=&\frac{GH^2(t_q)}{\pi^3}[\Gamma(\mu)]^22^{2\mu+5}\times\\&\left(\abs{C_q}^2+\abs{D_q}^2-2\Re(C_qD^{\star}_q)\right)
\end{split}
\end{eqnarray}
Substituting Eqs. (\ref{vacuumconditionIII}) and (\ref{normalizationIII}) into the above equation and considering terms linear in the slow-roll parameters, gives
\begin{eqnarray}\label{tensoralphavacuaDr.}
\begin{split}
^{\alpha}\mathcal{P}^{(T)}_{\mathcal{D},0}(t_q,q=\mathcal H(t_q))=&\frac{GH^2(t_q)}{\pi^2}[\Gamma(\mu)]^22^{2\mu+3}\times\\&\big[1+\frac{\left(2\epsilon(\tau_i)-\eta(\tau_i)\right)}{q\tau_i}\sin(2q\tau_i)\big]
\end{split}
\end{eqnarray}
This can be simplified using Eqs. (\ref{alphavacuaconditionII}), (\ref{initialtimeIV}), and (\ref{initialtimeVII}) as
\begin{eqnarray}\label{tensoralphavacuaIII}
\begin{split}
^{\alpha}\mathcal{P}^{(T)}_{\mathcal{D},0}(t_q,q=\mathcal H(t_q))=&\frac{GH^2_{l}}{\pi^2}[\Gamma(\mu)]^22^{2\mu+3}(\frac{q}{q_{l}})^{2\epsilon}\Big\{1+\frac{H_{f}}{\Lambda}\big[2\epsilon_{f}(\frac{q}{q_{f}})^{\epsilon+2\eta}-\\&\eta_{f}(\frac{q}{q_{f}})^{-2\eta+\xi}\big]\sin(\frac{2\Lambda(q/q_{f})^{\epsilon}}{H_{f}})\Big\}
\end{split}	
\end{eqnarray}
Following the same procedure that leads to Eq. (\ref{spectralindexI}), we can evaluate the tensor spectral index of Eq. (\ref{tensorspectralindexI}). The result is
\begin{eqnarray}\label{tensoralphavacuaIV}
\begin{split}
n_t(q)=&-2\epsilon_{l}(\frac{q}{q_{l}})^{-2\epsilon}+\\&2\epsilon_{f}\left(2\epsilon_{f}(\frac{q}{q_{f}})^{4\epsilon+4\eta}-\eta_{f}(\frac{q}{q_{f}})^{3\epsilon+\xi}\right)\cos(\frac{2\Lambda(q/q_{f})^{\epsilon}}{H_{f}})
\end{split}
\end{eqnarray}
We see that the leading correction of tensor spectral index due to $\alpha$-vacua is the same as that of scalar spectral index. In \cite{Broy:2016zik}, similar result has been obtained for de-Sitter inflation.
\section{Comparison with Observational Data}
\label{sec:Comparison with Observational Data}
In this section we are going to compare the trans-Planckian power spectrum (\ref{finalscalar}) with observational data and obtain constraints on parameters of it. Besides this,  it is noteworthy to explore another possibilities of power spectra which exhibit radical departure from the nearly scale invariant power spectrum, i.e. Eq. (\ref{powerspectrumDr.}) which is often written as $\mathcal{A}_s(q/q_{*})^{n_s-1}$ (with best-fit values for its parameters as $\mathcal{A}_s=2.215\times10^{-9}$ and $n_s=0.9624$ \cite{Ade:2015lrj}). 

One possibility is the logarithmic modulation of the primordial power spectrum
\begin{eqnarray}\label{logarithmic}
\mathcal{P}_{\mathcal{R}}^{log}(q)=\mathcal{P}_{\mathcal{R}}^{0}(q)\left\{1+\mathcal{A}_{log}\cos\left[\omega_{log}\ln(\frac{q}{q_{*}})+\phi_{log}\right]\right\}
\end{eqnarray}
Similar to our work, this kind of primordial power spectrum usually appears when non-Bunch-Davies initial condition is imposed for inflaton fluctuations \cite{Martin:2000xs,Danielsson:2002kx,Bozza:2003pr} or in the axion monodromy model \cite{Silverstein:2008sg,Kobayashi:2012kc,McAllister:2008hb}. The best-fit values for the free parameters of this model are $\mathcal{A}_{log}=0.0278$, $\log_{10}\omega_{log}=1.51$, and $\phi_{log}/2\pi=0.634$ \cite{Ade:2015lrj}. 

Boundary effective field theories \cite{Jackson:2013vka,Meerburg:2013dla} can lead to linear oscillation of the primordial power spectrum
\begin{eqnarray}\label{linear}
\mathcal{P}_{\mathcal{R}}^{lin}(q)=\mathcal{P}_{\mathcal{R}}^{0}(q)\left[1+\mathcal{A}_{lin}(\frac{q}{q_{*}})^{n_{lin}}\cos\left(\omega_{lin}\frac{q}{q_{*}}+\phi_{lin}\right)\right]
\end{eqnarray}
where the best-fit values for its free parameters are $\mathcal{A}_{lin}=0.0292$, $n_{lin}=0.662$, $\log_{10}\omega_{lin}=1.73$, and $\phi_{lin}/2\pi=0.554$ \cite{Ade:2015lrj}.

To compare these spectra with (\ref{finalscalar}), let us write it as follows
\begin{eqnarray}\label{observationI}
\begin{split}
&^{\alpha}\mathcal{P}^{(S)}_{\mathcal{R},0}(t_q,q=\mathcal H(t_q))=\mathcal{A}_{tp}(\frac{q}{q_l})^{n_{tp}-1}\times\\&\Big\{1+\gamma\big[2\epsilon_{f}(\frac{q}{q_{f}})^{\epsilon+2\eta}-\eta_{f}(\frac{q}{q_{f}})^{-2\eta+\xi}\big]\sin(\frac{2(q/q_{f})^{\epsilon}}{\gamma})\Big\}
\end{split}
\end{eqnarray}
where the subscript $tp$ in $n_{tp}$ stands for trans-Planckian and we have defined
\begin{eqnarray}\label{observationII}
\mathcal{A}_{tp}\equiv\frac{G}{\pi^2}\frac{H^2_{l}}{\epsilon_{l}}2^{2\nu-1}[\Gamma(\nu)]^2,~~~\gamma\equiv\frac{H_f}{\Lambda}
\end{eqnarray}
Throughout this section, we treat the pivot scales as constant, i.e. $q_{*}=q_l=0.05~1/Mpc$ and $q_f=0.01~1/Mpc$.

\begin{figure}[h!]
\begin{center}
\includegraphics[scale=0.7]{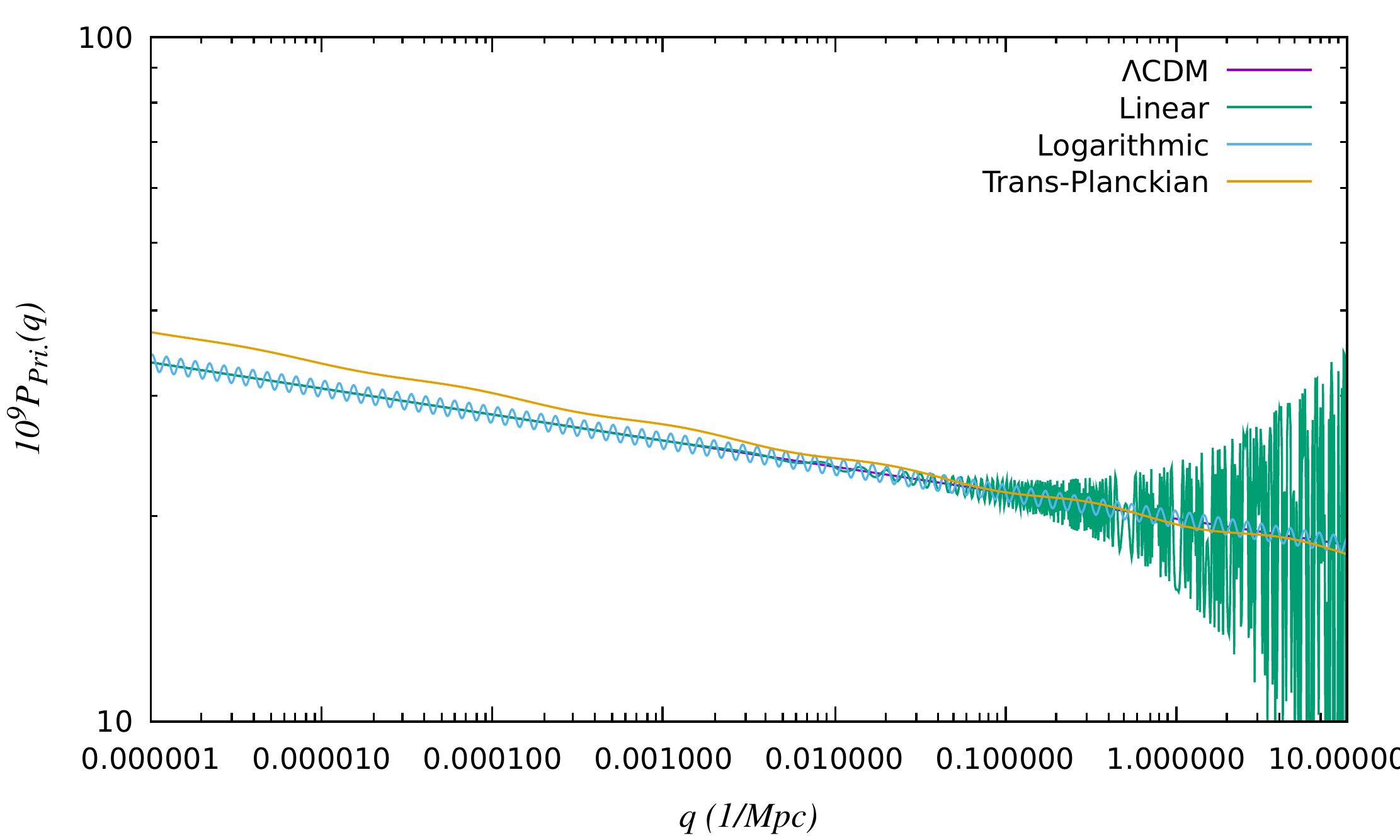}
\caption{Dimensionless primordial power spectra. Violet line is the $\Lambda$CDM primordial power spectrum, Eq. (\ref{powerspectrumDr.}). Green and blue lines are the linear and logarithmic primordial power spectra, Eq. (\ref{linear}) and Eq. (\ref{logarithmic}) respectively. These power spectra have been plotted by using the best-fit values of their parameters. Orange line is the primordial power spectrum in Eq. (\ref{observationI}) with best-fit values of its parameters from Table (\ref{tab:table2}).}
\label{fig:01}
\end{center}
\end{figure}

\begin{figure}[h!]
\includegraphics[scale=0.7]{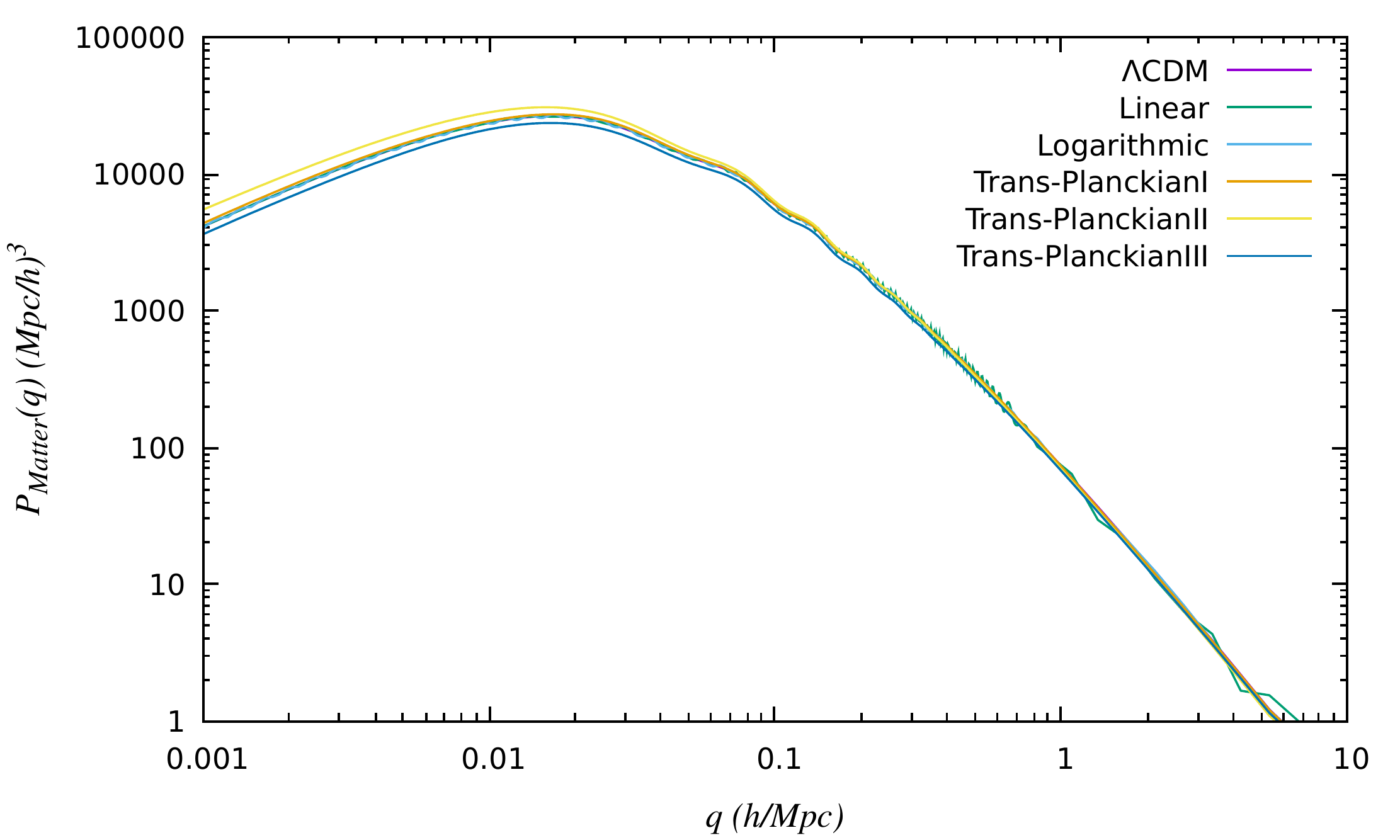}
\centering
\caption{The matter power spectra at $z=0$ for $\Lambda$CDM (violet), linear (green), and logarithmic (blue), using the best-fit values of their parameters. Orange line is the result for the trans-Planckian primordial power spectrum by using the best-fit values in Table (\ref{tab:table2}). Yellow line is the matter power spectrum for the trans-Planckian primordial power spectrum by using $\mathcal{A}_{tp}=2.5\times10^{-9}$, $n_{tp}=9.3\times10^{-1}$, and $\gamma=7\times10^{-2}$. Dark blue line is the result for trans-Planckian case with $\mathcal{A}_{tp}=2\times10^{-9}$, $n_{tp}=9.7\times10^{-1}$, and $\gamma=7\times10^{-3}$. The rest of the parameters in the last two plots are those in Table (\ref{tab:table2}).}
\centering
\label{fig:02}
\end{figure}

\begin{figure}[h!]
\includegraphics[scale=0.7]{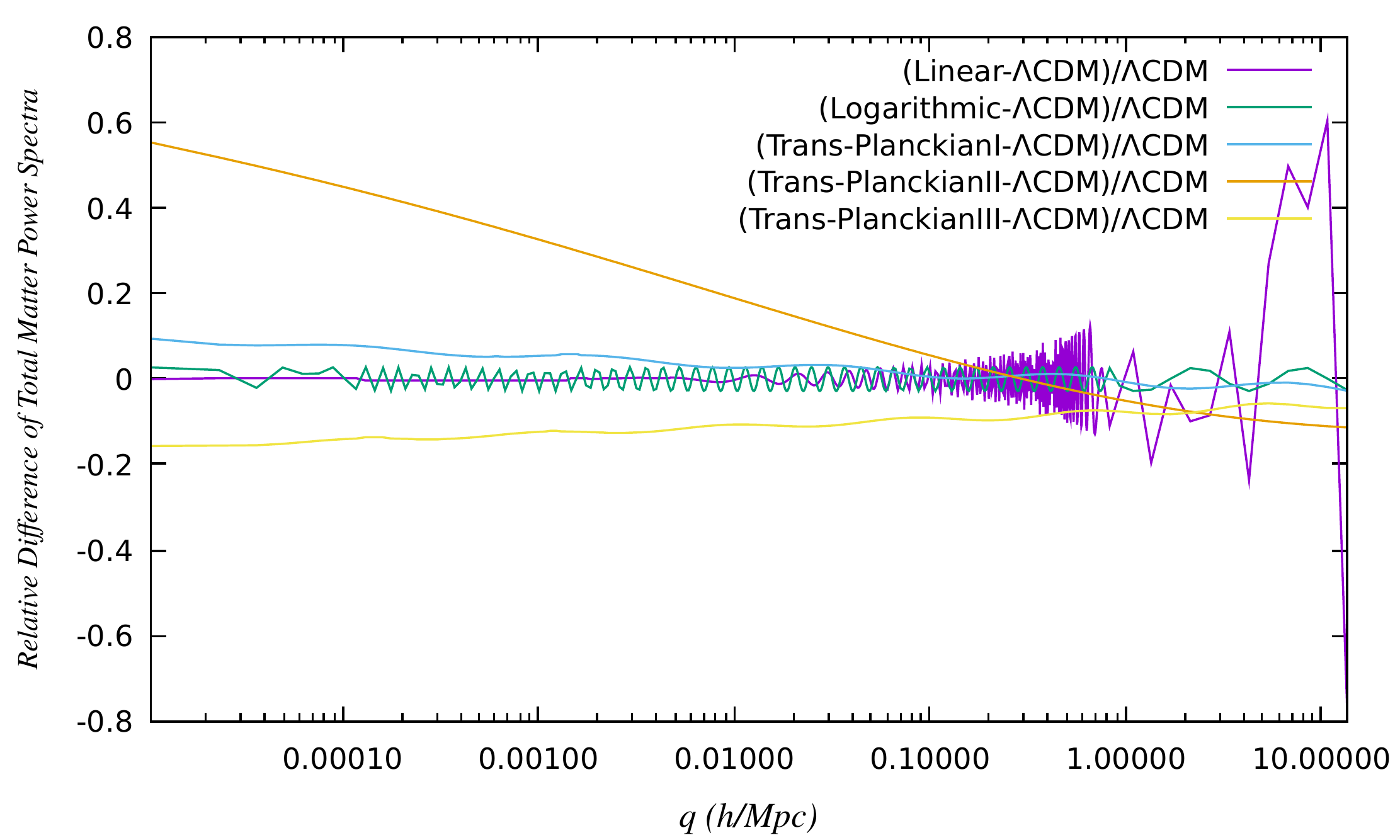}
\centering
\caption{Relative difference of the matter power spectra with $\Lambda$CDM model. Violet and green lines are the relative difference of the linear and logarithmic matter power spectra with that of $\Lambda$CDM model by using the best-fit values of their parameters. Blue line is this relative difference for the trans-Planckian model with the best-fit values of the parameters in Table (\ref{tab:table2}). Orange line is this relative difference for trans-Planckian model with $\mathcal{A}_{tp}=2.5\times10^{-9}$, $n_{tp}=9.3\times10^{-1}$, and $\gamma=7\times10^{-2}$. Yellow line is this relative difference by using $\mathcal{A}_{tp}=2\times10^{-9}$, $n_{tp}=9.7\times10^{-1}$, and $\gamma=7\times10^{-3}$. The rest of the parameters in these last two plots are those in Table (\ref{tab:table2}).}
\centering
\label{fig:03}
\end{figure}

\begin{figure}[h!]
\includegraphics[scale=0.7]{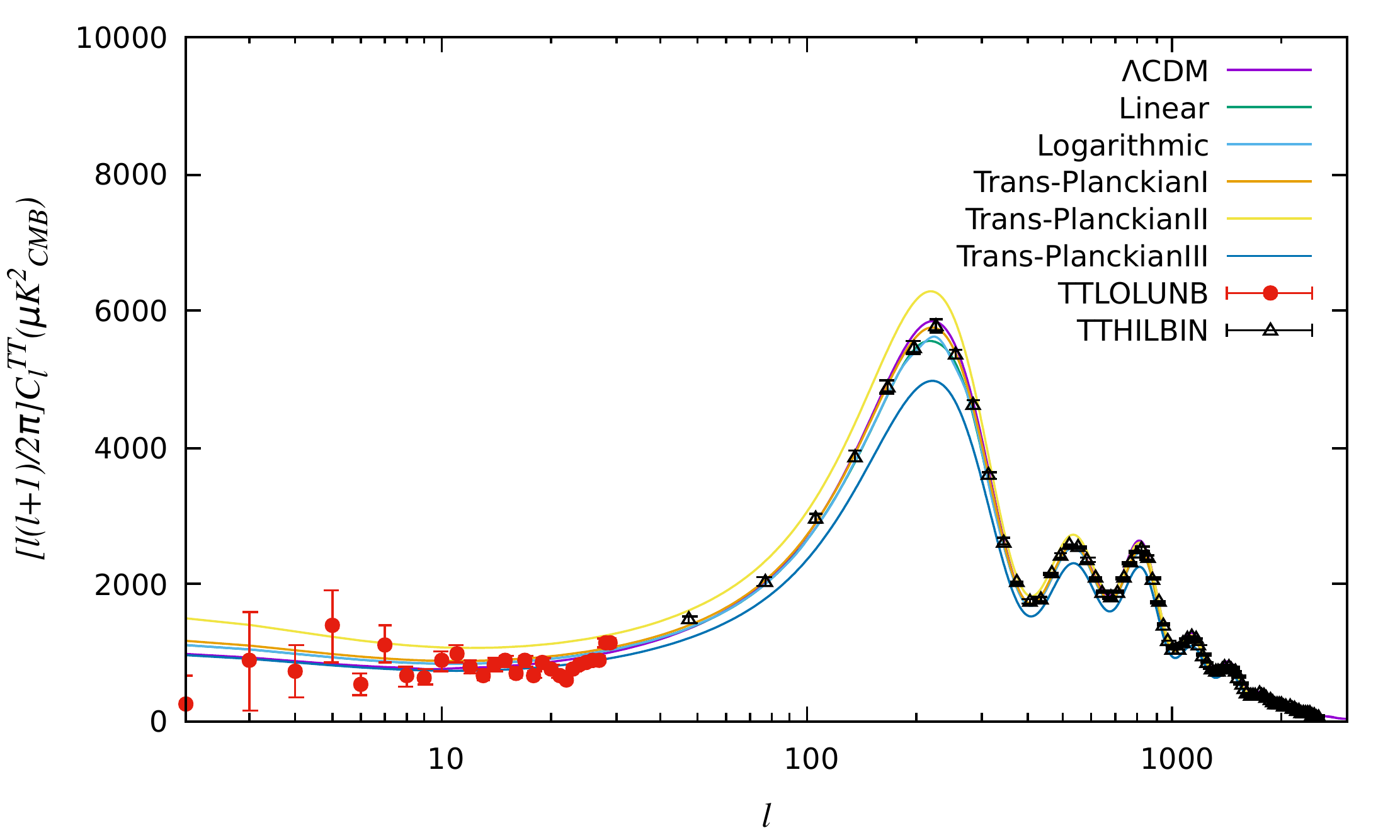}
\centering
\caption{CMB temperature anisotropy. Violet, green, and blue lines are the results for $\Lambda$CDM, linear, and logarithmic, respectively. These results have been plotted by using the best-fit values of parameters. Orange line is the trans-Planckian result with the parameters in Table (\ref{tab:table2}). Yellow line is the trans-Planckian result by using $\mathcal{A}_{tp}=2.5\times10^{-9}$, $n_{tp}=9.3\times10^{-1}$, and $\gamma=7\times10^{-2}$. Dark blue line is the result for trans-Planckian case with $\mathcal{A}_{tp}=2\times10^{-9}$, $n_{tp}=9.7\times10^{-1}$, and $\gamma=7\times10^{-3}$. The rest of the parameters in these last two plots are those in Table (\ref{tab:table2}). Red dots are the Planck low $l$ unbinned results (TTLOLUNB). Black dots are the Planck high $l$ binned results (TTHILBIN). The Planck data are plotted at $1\sigma$ confidence level.}
\centering
\label{fig:04}
\end{figure}

In order to do the comparison, we use the CLASS (Cosmic Linear Anisotropy Solving System) code \cite{Blas:2011rf} alongside with the Monte Python \cite{Audren:2012wb}. Also the Nested Sampling method (through MultiNest \cite{Feroz:2008xx}) is used. The best-fit values of the trans-Planckian power spectrum, Eq. (\ref{observationI}), are given in Table (\ref{tab:table2}). In Fig. (\ref{fig:01}), we have plotted the $\Lambda$CDM (Eq. (\ref{powerspectrumDr.})), trans-Planckian (Eq. (\ref{observationI})), logarithmic (Eq. (\ref{logarithmic})), and linear (Eq. (\ref{linear})) primordial power spectra.
\renewcommand{\arraystretch}{2}
\begin{table}[h!]
\begin{center}
\caption{Best-fit values for the trans-Planckian power spectrum in Eq. (\ref{observationI}).}
\label{tab:table2}
\begin{tabular}{l|c|c|c|c} 
\hline 
Param & best-fit & mean$\pm\sigma$ & 95\% lower & 95\% upper \\ \hline
$10^{+9}A_{s }$ &$2.251$ & $2.245_{-0.0034}^{+0.0034}$ & $2.238$ & $2.252$ \\
$n_{s }$ &$0.9536$ & $0.9479_{-0.0021}^{+0.0022}$ & $0.9435$ & $0.9523$ \\
$\epsilon$ &$0.01026$ & $0.08872_{-0.015}^{+0.017}$ & $0.04313$ & $0.1274$ \\
$\eta$ &$0.02805$ & $0.03562_{-0.036}^{+0.0088}$ & $3.145\times10^{-7}$ & $0.08005$ \\
$\xi$ &$0.04818$ & $0.09979_{-0.032}^{+0.028}$ & $0.03523$ & $0.1697$ \\
$\gamma$ &$0.0091$ & $0.007936_{-0.00014}^{+0.0021}$ & $0.002639$ & $0.01$ \\
$\epsilon_{f }$ &$0.4842$ & $0.5007_{-0.03}^{+0.028}$ & $0.439$ & $0.5676$ \\
$\eta_{f }$ &$0.05039$ & $0.1003_{-0.029}^{+0.03}$ & $0.03202$ & $0.1648$ \\
\end{tabular} \\
\end{center}
\end{table}\\
\renewcommand{\arraystretch}{1}
\\
One important quantity in describing the structure formation of the Universe is the matter power spectrum, which is the difference between the local density and the mean density as a function of scale. It is the Fourier transform of the matter correlation function and can be expressed as follows \cite{Dodelson:2003ft}
\begin{eqnarray}\label{totalmatterpowerspectrum}
P_{Matter}(q,z)=\frac{4}{25}\frac{q^4T^2(q)D^2(z)}{H_0^4\Omega^2_{Matter}}\mathcal{P}_{\mathcal{R}}(q)
\end{eqnarray}
where $z$ is the redshift, $T(q)$ is the transfer function, $D(z)$ is the growth function, $H_0$ is the present value of Hubble parameter, $\Omega_{Matter}$ is the matter density parameter, and $\mathcal{P}_{\mathcal{R}}(q)$ is the primordial power spectrum.
In Fig. (\ref{fig:02}) we have plotted the matter power spectrum at redshift $z=0$ for $\Lambda$CDM, linear, logarithmic, and trans-Planckian primordial power spectra. For further insight, the relative difference of the matter power spectra with the $\Lambda$CDM matter power spectrum are shown in Fig. (\ref{fig:03}). For instance, the blue line is the matter power spectrum for trans-Planckian model (with the best-fit values of its parameters in Table (\ref{tab:table2})) minus the matter power spectrum for $\Lambda$CDM model divided by the matter power spectrum for $\Lambda$CDM model. From this figure, we infer that at small values of $q$, the matter power spectrum for the trans-Planckian model is bigger than that of the $\Lambda$CDM model. By increasing the wave number, this relative difference of the matter power spectra goes to zero with an oscillating behavior, and therefore there is no difference between these two models at larger values of $q$. Orange line is this relative difference when $\mathcal{A}_{tp}=2.5\times10^{-9}$, $n_{tp}=9.3\times10^{-1}$, and $\gamma=7\times10^{-2}$. Yellow line is the relative difference but this time for $\mathcal{A}_{tp}=2\times10^{-9}$, $n_{tp}=9.7\times10^{-1}$, and $\gamma=7\times10^{-3}$. Once again, these last two plots confirm that at larger values of the wave number, the discrepancy between the trans-Planckian and $\Lambda$CDM models becomes insignificant, regardless of the choice of the free parameters of the trans-Planckian primordial power spectrum.
\\
\\
Different primordial power spectra, can leave imprints on the CMB temperature anisotropy. In Fig. (\ref{fig:04}), by using the best-fit values of the free parameters, we have plotted the lensed CMB temperature anisotropy for $\Lambda$CDM (violet), linear (green), logarithmic (blue), and trans-Planckian (orange) models in comparison with the Planck data. Comparison of these four plots by the Planck data, we conclude that these types of the primordial power spectra, have no significant difference in the CMB temperature anisotropy. However, a small departure between these models can be seen for intermediate values of $l$. Yellow line is the trans-Planckian result by choosing $\mathcal{A}_{tp}=2.5\times10^{-9}$, $n_{tp}=9.3\times10^{-1}$, and $\gamma=7\times10^{-2}$. Dark blue line is the trans-Planckian result for $\mathcal{A}_{tp}=2\times10^{-9}$, $n_{tp}=9.7\times10^{-1}$, and $\gamma=7\times10^{-3}$. We see that this last plot has better adaptability with the Planck data for intermediate values of $l$.
\\

\section{Concluding Remarks}
\label{sec:Conclusion}
It is believed that the inflationary epoch erases any memories of the pre-inflationary initial state. This means that if the stage of inflation continues for a sufficiently long time, all information about the initial conditions including any physically reasonable choice of the vacuum state would be erased. But this is not correct if the inflation does not last too long or in the presence of perturbations modes' quanta with very high initial momenta.\footnote{The latter had already been raised in Hawking radiation of black holes. At asymptotic infinity, the observed photons in the high energy tail of the black body radiation are red shifted from a trans-Planckian value. It is interesting that in this context, thermalization plays the role of inflation in cosmology. That means that the thermalization process erases the effects of non-adiabatic evolution \cite{bl}.}In both cases, there would be some imprint on the late time observables. This gives the motivation to consider some modified initial states of the universe different from Bunch-Davis type. On the other hand if the inflation lasts more than 60 e-folds, some of the physical scales observed in the CMB spectrum, were in the super Planckian regime at the onset of inflation. The properties of these scales are sensitive to the conditions provided at that time. In the absence of a theory of quantum gravity, the effects of trans-Planckian era usually parametrized by modified dispersion relations or by modified vacuum states, $\alpha$ vacua.

Here a question maybe raised. Is any feature of the trans-Planckian physics realized in the observation. To answer this question \cite{intd, aa,jcap}, it must be noted that the observation of anisotropies of CMB leads to some indirect information about quantities evolved from an initial state. The relation of these quantities and the CMB spectrum dependes to extra parameters such as the density of dark and baryonic matter, the Hubble constant etc. which are not related to inflation. Thus any correction to the spectrum of CMB may be related to these parameters and not necessarily to the effects of trans-Planckian physics. The existence of  some noisy sources that encode inhomogeneities, the volume-averaged observations,  accessibility a finite range of momenta etc. can affect CMB spectrum.

Going over to the pre-inflationary initial conditions that differ from the usual Bunch-Davies vacuum, here we have used Danielsson's $\alpha$-vacua prescription in the slow-roll regime. The motivation for this kind of vacuum is that the duration of inflation is finite and therefore the Minkowskian vacuum in infinite past may not be accessible. Therefore, one should impose the initial condition at a finite time, say $t_i$ or equivalently $\tau_i$. It is useful to remember that dealing with the Bunch-Davies vacuum, it is convenient to evaluate different quantities at the time of horizon crossing, $t_q$, which is $q$-dependent. This motivates us to define one pivot scale, $q_{l}$, which is the last wave number that leaves the horizon (see Eq. (\ref{q-dependenceIII})). As we have shown by Eq. (\ref{alphavacuaconditionII}), in the case of $\alpha$-vacua, the initial time is also $q$-dependent. Thus, in addition to  $q_{l}$, we have introduced in Eq. (\ref{Hubbleatfirstscale}) the second pivot scale, $q_{f}$, which is the first wave number that satisfies the initial condition.\\
Using these definitions, we have calculated corrections to the scalar and tensor power spectra and comparison with the results of standard Bunch-Davies  has been done. Also, the scalar and tensor spectral indices have been computed and we have found that the leading correction to these quantities has a $q$-dependent amplitude, $2\epsilon_{f}[2\epsilon_{f}(\frac{q}{q_{f}})^{4\epsilon+4\eta}-\eta_{f}(\frac{q}{q_{f}})^{3\epsilon+\xi}]$, which is obviously second order in slow-roll parameters when are evaluated at $q=q_{f}$. Perhaps, the exquisite feature of our work is that the leading correction to $n_s(q)$ and $n_t(q)$ has an oscillatory part, $\cos(\frac{2\Lambda(q/q_{f})^{\epsilon}}{H_{f}})$, which is also $q$-dependent.\\
As previously mentioned, the idea of imposing the initial condition at a finite time was used for the first time by Danielsson \cite{Danielsson:2002kx}. For purely de Sitter inflation in which the Hubble parameter is constant, he obtained the following result for the scalar power spectrum (see equation (32) in \cite{Danielsson:2002kx})
\begin{eqnarray}\label{Danielsson'sresult}
\mathcal{P}^{(S)}_{\mathcal{R}}(t_q,q=\mathcal{H}(t_q))=(\frac{H}{2\pi})^2[1-\frac{H}{\Lambda}\sin(\frac{2\Lambda}{H})]
\end{eqnarray}
which has constant amplitude and frequency. Most recently, similar calculation has been done for quasi-de Sitter inflation by Broy \cite{Broy:2016zik}. In his paper, the mode functions have been obtained by assumption that the Hubble parameter is truly constant. Then some corrections, similar to what is obtained by Danielsson \cite{Danielsson:2002kx}, for quasi-de Sitter power spectrum have been estimated. And in calculating the spectral indices, the Hubble and slow-roll parameters are considered to be $q$-dependent. To be more precise, in our present work, we have extended the results of  \cite{Danielsson:2002kx} and \cite{Broy:2016zik} to the slow-roll inflation. According to our straightforward calculations, Eqs. (\ref{finalscalar}) and (\ref{tensoralphavacuaIII}), the scalar and tensor power spectra are $q$-dependent through $t_q$ and $t_i$.
\section*{Acknowledgments}
H. Bouzari Nezhad would like to thank R. Moti for some fruitful disscusion. This work has been supported by a grant from university of Tehran.

\end{document}